# Direct Opto-Electronic Imaging of 2D Semiconductor- 3D Metal Buried Interfaces

Kiyoung Jo[1], Pawan Kumar[1,2], Joseph Orr[3], Surendra B. Anantharaman[1], Jinshui Miao[1], Michael Motala[4,7], Arkamita Bandyopadhyay[2], Kim Kisslinger[5], Christopher Muratore[6], Vivek B. Shenoy[2], Eric Stach[2], Nicholas Glavin[4], Deep Jariwala[1]*

[1]Electrical and Systems Engineering, University of Pennsylvania, Philadelphia, PA, 19104, USA

[2]Materials Science and Engineering, University of Pennsylvania, Philadelphia, PA, 19104, USA

[3]Electrical and Computer Engineering, Villanova University, Villanova, PA, 19085, USA

[4]Air Force Research Laboratory, Materials and Manufacturing Directorate, Wright-Patterson AFB, Columbus, OH, 45433, USA

[5]Brookhaven National Laboratory, Upton, NY, 11973, USA

[6]University of Dayton, Dayton, OH, 45469, USA

[7]UES Inc., Beavercreek, OH, 45432, USA

Abstract

The semiconductor-metal junction is one of the most critical factors for high performance electronic devices. In two-dimensional (2D) semiconductor devices, minimizing the voltage drop at this junction is particularly challenging and important. Despite numerous studies concerning contact resistance in 2D semiconductors, the exact nature of the buried interface under a three-dimensional (3D) metal remains unclear. Herein, we report the direct measurement of electrical and optical responses of 2D semiconductor-metal buried interfaces using a recently developed metal-assisted transfer technique to expose the buried interface which is then directly investigated using scanning probe techniques. We characterize the spatially varying electronic and optical properties of this buried interface with < 20 nm resolution. To be specific, potential, conductance and photoluminescence at the buried metal/$MoS_2$ interface are correlated as a function of a variety of metal deposition conditions as well as the type of metal contacts. We observe that direct evaporation of Au on $MoS_2$ induces a large strain of ~5% in the $MoS_2$ which, coupled with charge transfer, leads to degenerate doping of the $MoS_2$ underneath the contact. These factors lead to improvement of contact resistance to record values of 138 k$\Omega$ μm, as measured using local conductance probes. This approach was adopted to characterize $MoS_2$-In/Au alloy interfaces, demonstrating contact resistance as low as 63 k$\Omega$ μm. Our results highlight that the $MoS_2$/Metal interface is sensitive to device fabrication methods, and provides a universal strategy to characterize buried contact interfaces involving 2D semiconductors.

## Keywords

$MoS_2$; Two-dimensional Semiconductor - Metal Interface; Buried Interface; Kelvin probe force Microscopy; Tip-enhanced Raman spectroscopy

# Introduction

The semiconductor-metal junction is one of the most consequential for electronic and optoelectronic device performance. This junction has been a crucial bottle neck for improvement of nano-electronic device performance and minimizing power consumption. For < 20 nm channel length ballistic field effect transistors, the voltage drop due to metal-semiconductor contact resistance is the primary source of energy dissipation and also contributes to parasitic circuit elements, thereby limiting high frequency operation. Metal contacts for semiconductors are often imperfect due to surface effects such as Fermi level pinning, incomplete passivation of surface dangling bonds and chemical reactivity of the metal with the semiconductor, all of which have long plagued many 3D bulk semiconductors such as Si, Ge, and the III-V family.[1–3] Traditionally, a major obstacle to understanding contact resistance has been the buried nature of the metal-semiconductor interface. This interface is confined by both the semiconductor and metal, and it is therefore impossible to directly measure electronic and optical signals, and map the 2D interface.

The advent of atomically-thin two-dimensional (2D) semiconductors has renewed enthusiasm and interest in post-silicon channel candidates given their superior electrostatic control. In this context, the 3D metal contacts to 2D semiconducting channels are of paramount importance. However, the utility of perfect contacts stems far beyond just transistors, but also to photovoltaics, LEDs and other opto-electronic devices.[4,5] Unlike 3D semiconductor surfaces, however, most 2D semiconductors (and the chalcogenides in particular) have self-passivated, oxide free surfaces. While this avoids the issue of unpassivated dangling bonds and surface oxides on the semiconductor side, it creates additional challenges in providing an electronically intimate, uniform and low-resistance contact with 3D metals, even when the contact metal is an inert noble metal such as Au. A large body of work on bulk 3D metal/2D semiconductor contacts is available.[6–8] A number of metals with varying work functions and compositions have been investigated as electrical contacts to 2D semiconductors, namely transition metal dichalcogenides (TMDCs) ($MX_2$ ; M= Mo, W; X = S, Se, Te), in the context of field-effect devices such as transistors.[7,8] In addition, the impact of deposition parameters and methods has also been thoroughly investigated.[8] It was found that oxide forming metals tend to chemically react with 2D semiconductors forming unstable contacts with contact resistance increases or becomes non-ohmic with time.[6,8,9] On the other hand, use of high energy or lower pressure deposition methods such as e-beam evaporation or ion beam sputtering results in bombardment-induced damage, creating interface states that result in Fermi level pinning.[9,10] In such cases, the use of mechanical stamping[10] or low-melting point metals such as Indium[11] and its alloys has resulted in low-resistance contacts with pristine interfaces. Surface chemistry and phase engineering has also been employed in TMDCs on Mo and W which induces 2H (semiconducting) to 1T' or 1T phase transition that reduces the contact resistance with bulk 3D metals.[12–14] This phase transition can be induced via lithium intercalation,[13] local heating,[15] or even using local electron beam irradiation[16]. Thus, significant research has been performed on making high quality contacts between 3D metals and 2D semiconductors.

In terms of characterizing this 3D/2D metal/semiconductor buried interface, prior research has largely focused on localized structure measurement via cross-sectional transmission electron microscopy (TEM)[10] or Raman spectroscopy of evaporated metal nano islands[17,18] on the 2D semiconductor. Several studies on current-voltage responses on field-effect devices[10] have also been conducted but such measurements are indirect and ensemble in nature.

Scanning probe micro-spectroscopy with a tip-based local probe is a powerful technique for comprehensive nanoscale characterization as it can provide optical, electrical and structural information of the surface with high spatial resolution and is only limited by the tip sharpness or diameter.[19] The tip-based local probe not only allows mapping of local topography and electrical properties, but by taking advantage of the ultra-sharp apex of the metal tips, it also allows concurrent probing of optical properties via the plasmonic gap mode that is formed between the tip and a plasmonic substrate. The gap-mode further helps enhance optical signals by virtue of the high local electric fields and the Purcell effect for photoluminescence and Raman scattering phenomena. These combined features enable a multi-functional, non-destructive, and versatile technique to probe opto-electronic interfaces. Several studies on 2D TMDCs using such scanning local-probe micro-spectroscopies have been already performed, including correlation of optical and electronic properties at crystal defects.[20,21] However, to the best of our knowledge this multi-functional technique has not been used to characterize and understand the buried metal-semiconductor interface.

In this study, we investigate the $MoS_2$- metal buried interface directly via the tip-based scanning probe technique. By correlating electronic techniques such as Kelvin Probe Force Microscopy (KPFM), Conductive AFM (C-AFM) with optical information such as tip-enhanced Raman (TERS) and photoluminescence spectroscopy (TEPL), we have comprehensively evaluated electronic, optical and mechanical properties of the buried interface, all concurrently measured with high spatial resolution (tens of nanometers). Our studies have also provided insight into $MoS_2$-metal interfaces formed via various metal deposition techniques as well as different metals. Our results highlight that metal contacts formed via evaporation are distinct as compared to van der Waals contacts and that In-Au based evaporated contacts offer the lowest resistance concurrently with high spatial homogeneity.

Results and Discussion

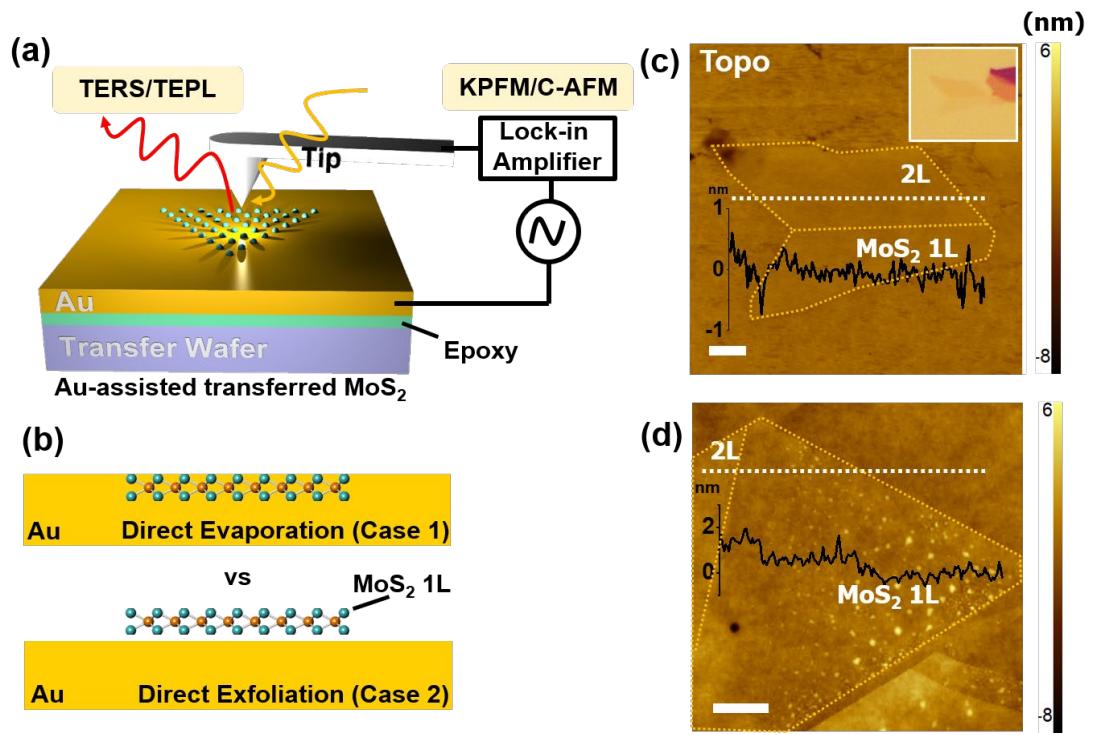

Figure 1. (a) Schematic representation of scanning probe techniques used in this study including KPFM, C-AFM, Tip-enhanced Raman and PL spectroscopy (b) Cross-sectional view of Au-assisted transfer of monolayer (1L) MoS$_2$ (top, referred to as Case 1) for directly evaporated metal contacts and directly exfoliated MoS$_2$ on Au (bottom, Case 2) that are investigated in this study (c, d) topography map of (c) Case 1 and (d) Case 2. Inset at (c) is an optical image of the MoS$_2$ flake transferred after direct Au evaporation. Graphs inside (c) and (d) represent the lateral height profile along dotted lines. Scale bars indicate 1 μm.

To investigate optical and electrical properties of 2D semiconductor-metal Interface, MoS$_2$ was used as a representative 2D semiconductor. To correlate optical and electrical properties, the scanning probe technique was used (figure 1a). Specifically, TERS and TEPL spectrum were obtained. In addition, KPFM and C-AFM were performed as electrical characterization on the same flake to avoid possible sample-to-sample variation. To characterize and compare the properties of two different types of interfaces, samples were prepared, (1) by direct evaporation of Au on $MoS_2$ followed by epoxy-assisted template stripping to expose the buried interface and (2) by direct mechanical exfoliation of $MoS_2$ on freshly template stripped Au films as shown in figure 1b. The electron beam evaporation technique was used to deposit Au.[22] Thermal evaporation and sputtering were also investigated (see supporting information S8 for details). Due to the high kinetic energy of incoming Au atoms during evaporation, Au atoms are likely to strongly couple with S atoms on the $MoS_2$ surface, leading to an intimate electronic contact with negligible tunneling or Schottky barriers. On the other hand, it is widely accepted that the 2D TMDC semiconductor-metal vdW interface consists of tunneling and Schottky barriers because $MoS_2$ only forms a physical junction with Au without chemical bond formation.[23] Therefore, these two representative samples serve as a good comparison to understand the TMDC-metal interface. Topography analysis of Case 1 (the interface with evaporated gold) clearly reveals a flat height profile of the $MoS_2$ flake, indistinguishable from surrounding Au with minimal surface roughness (0.188 nm rms on flake *vs* 0.330 off flake). For Case 2 (direct exfoliation), the $MoS_2$ sits atop of the Au and hence, shows clear atomic steps and thickness variance with significant roughness (0.734 nm on flake *vs* 0.368 off flake) in AFM topography. These AFM topography results suggest that direct metal evaporation offers more a more spatially uniform contact.

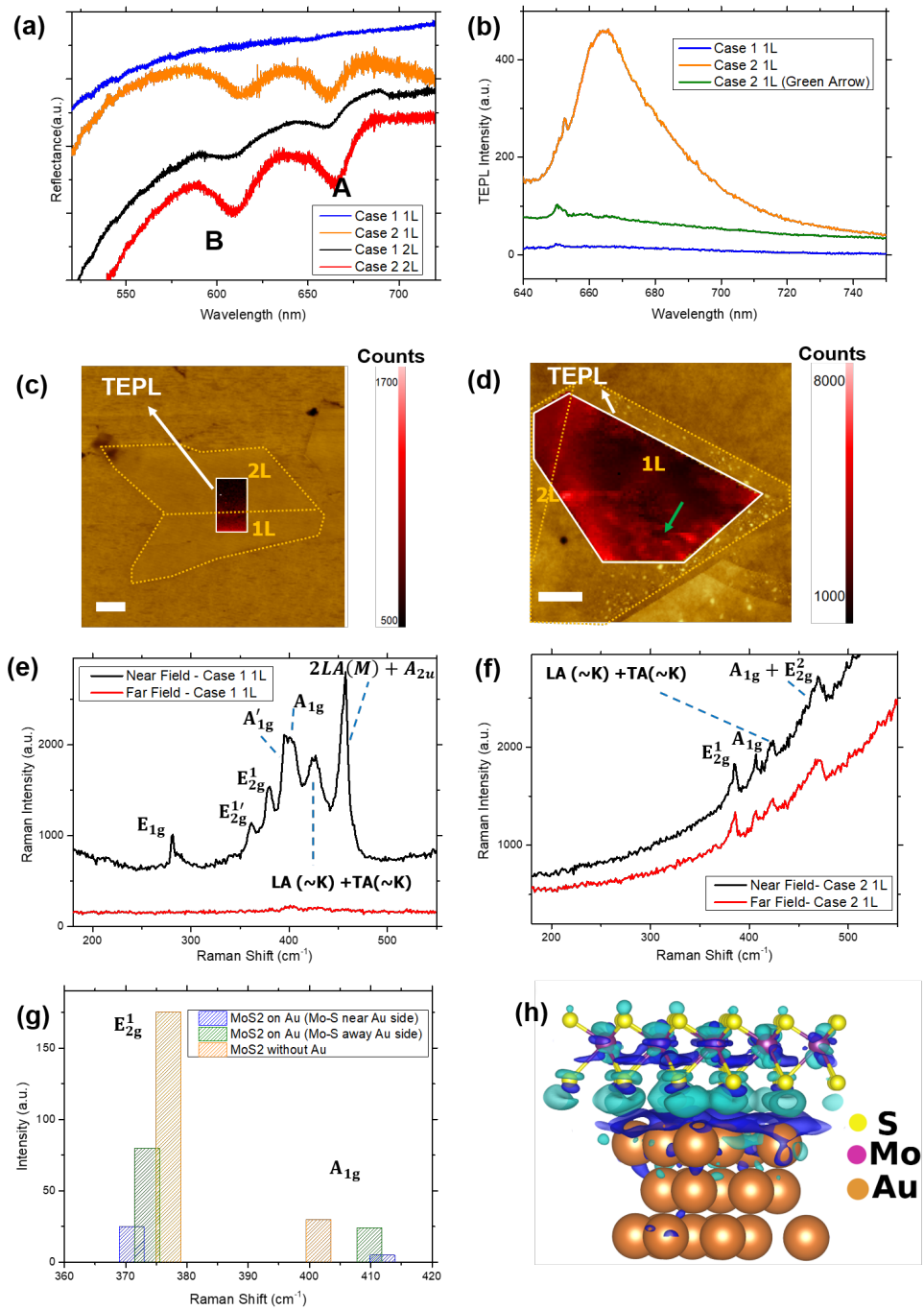

Figure 2. (a) Reflectance spectra of Case 1 and 2 with either monolayer (1L) or bilayer (2L) $MoS_2$. (b) Tip-enhanced photoluminescence (TEPL) spectrum of Case 1 and 2 with monolayer $MoS_2$. The yellow and green spectra are from two separate points on the directly exfoliated $MoS_2$. The green spectrum was collected at the location indicated by the green arrow shown in (d) at a quenched region. (c,d) TEPL map of (c) Case 1 and (d) Case 2. Scale bars indicate 1 μm. (e, f) Near-field (black) and far-field (red) Raman spectrum of (e) Case 1 and

(f) Case 2 with monolayer $MoS_2$. (g) First principles calculation of peak intensities of the two characteristic Raman modes of $MoS_2$ with and without Au layer. The distinct blue and green bars result from asymmetric Mo-S bond length change due to single-sided contact of $MoS_2$ with Au layer in the calculated model. (h) Charge density difference map between the $MoS_2$ and Au (111) surface. Dark blue and light blue stands for loss and gain of charge density, respectively.

The presence of strongly bound excitonic resonances in 2D TMDCs also provide an additional probe of the surrounding dielectric medium. In a metal-2D $MoS_2$ contact this can therefore be probed via far-field reflection spectroscopy. Figure 2a shows representative reflectance measurement of each contact. The vdW interface contact to monolayer and bilayer $MoS_2$ (Case 2) clearly shows two reflectance dips corresponding to A and B excitonic absorption at ~660 nm and ~610 nm, respectively.[24,25] On the contrary, the excitonic peaks were significantly suppressed in the direct Au evaporated sample (Case 1). For the case of monolayer $MoS_2$ with direct Au evaporation there is no clear evidence of excitonic resonances. This is particularly interesting because it has been reported that weak excitonic absorption peaks arise even in nanocrystalline $MoS_2$.[26] The complete disappearance of the excitonic transitions from a single crystalline exfoliated flake suggests that the degenerate doping via the evaporated metal contact is strong enough to completely screen out exciton formation by virtue of Pauli blocking. This further suggests that the monolayer $MoS_2$ surface is metallized or degenerately doped by the bottom Au contact. This result also shows that the Au based-transfer technique is powerful for preparation of metallized TMDC layers. For the bilayer sample however, the excitonic peaks reappear at ~ 661 and ~ 607 nm respectively for the A and B exciton. This further suggests that the metal induced degenerate doping is a strong function of separation from the metal and therefore the second layer in 2L sample shows evidence of excitons. However, the peaks are broader and weaker compared to directly exfoliated samples, which is an indication of significant charge transfer from the metal.

The scanning probe tip can also be effectively used as an antenna to obtain reliable, tip-enhanced Raman and photoluminescence signals[27,28] from these MoS$_2$-Au samples. Good electrical contact of a metal with a 2D semiconductor such as MoS$_2$ should result in degenerate doping that would result in photoluminescence suppression since high carrier density drives non-radiative recombination of excitons .[29]

TEPL was therefore conducted to understand excitonic emission for each MoS$_2$-Au interface. Overall, the photoluminescence (PL) intensity from MoS$_2$ for the MoS$_2$/directly evaporated Au interface is largely suppressed compared to that of the vdW interface (figure 2b, c, d). This clearly demonstrates that the buried interface formed via direct metal evaporation makes an electrically good contact in agreement with the reflectance results. On the other hand, the vdW contact interface samples emits high PL intensity at 665 nm which is the optical bandgap of MoS$_2$. This implies that the physical proximity with metal does not effectively quench the excitons in MoS$_2$ since the interface contains a small vdW gap. Further, it is clearly evident that the PL is spatially varying. To be specific, this large spatial inhomogeneity in PL is indicative of spatial inhomogeneity in the electronic quality of MoS$_2$-Au contact with certain "hotspot" regions (indicated by green arrow at figure 2b) of ~500 nm in lateral size showing reduced PL, suggesting that these "hotspot" regions are in relatively good contact with the Au.

Raman spectroscopy is another powerful tool to characterize 2D crystals including semi-quantitative analysis of carrier density and strain.[18,30–32] Coupled with a tip-based scanning probe technique, tip-enhanced Raman spectroscopy (TERS) can resolve Raman scattering with high spatial resolution. We investigated TERS for the direct evaporation and directly exfoliated

MoS$_2$-Au contact interfaces (figure 2e, f).

The Purcell effect combined with resonance Raman effect induces the expected remarkable increase in Raman scattering on the Case 1 interface monolayer sample and giving rise to multiple peaks as shown in figure 2e. The second-order zone-edge phonon 2LA(M) mode and first-order optical phonon A$_{2u}$ mode arising at ~456 cm$^{-1}$ were observed, and were due to plasmonic enhancement.[18] In-plane mode E$_{1g}$, which is normally forbidden in backscattering Raman process, also arises in this system as previously reported.[32,33] More importantly, strain induced peak splitting of $E_{2g}^1$ and $A_{1g}$ modes appear for MoS$_2$ with directly evaporated Au contacts.[18,34] The presence of peak splitting is attributed to change of symmetry from D$_{3h}$ to C$_s$ for the monolayer and from D$_{6h}$ to C$_{2h}$ for the bilayer.[18] A linear relationship of Raman peak position was reported to demonstrate a shift of 4.5 cm$^{-1}$/% strain for $E_{2g}^{1\prime}$ mode and 1.0 cm$^{-1}$/% strain for $A_{1g}$ mode.[34] The two shifting rates were well matched with the shift of $E_{2g}^{1\prime}$ and $E_{2g}^1$ modes obtained in this work, both of which suggest a ~5.5% of strain on the surface. This is considerably higher than other reports of ~2.0% for an Ag dendrite-MoS$_2$ interface, 3.0% for MoS$_2$-Au nano island, and ~1.7% for Ag coated CVD grown MoS$_2$.[18,33,35] These differences can be attributed to many factors such as different metal layer compositions, metal morphology, and the corresponding contact area between the metal and semiconductor. This large strain can be attributed to the lattice mismatch between the Au and MoS$_2$. Interestingly, the monolayer MoS$_2$ Case 1 interface did not exhibit clear Raman peaks under far-field Raman measurements which were collected by retracting the tip away by ~30 nm from the samples. The Raman intensity suppression is likely due to intimate Au contact which holds MoS$_2$ atoms tightly hindering vibration for each atom. Besides, intimate Au contact induces high carrier concentration on MoS$_2$ leading to the Raman suppression as reported in previous gate-dependent far field Raman studies.[31] The Raman spectra for the Case 1 interface further suggests that there is no electronic or structural transformation in the MoS$_2$ upon metal evaporation, and the contact formation only induces strong degenerate doping and strain in the MoS$_2$ layer. On the other hand, the Raman spectrum of monolayer MoS$_2$ Case 2 interface showed negligible strain, maintaining the unstrained $E_{2g}^1$ peak position at 385 cm$^{-1}$.[36] It implies that MoS$_2$ and Au do not strongly couple, but physisorption governs the contact which does not impose significant strain measurable via Raman. This stark contrast clearly shows the Case 1 interface is fundamentally different from the Case 2 interface in terms of strain.

Another independent factor which contributes to the electrical contact at a metal-semiconductor interface is charge transfer. Semiconductor band bending will result in the accumulation of surface charge. This charge can be qualitatively analyzed by the A$_{1g}$ mode which is the out-of-plane Raman mode strongly affected by electron-phonon interaction and results in softening and broadening at high electron density in MoS$_2$ for resonant Raman.[30,31] This is because high carrier density leads to Pauli blocking at the K point in the conduction band of MoS$_2$, resulting in reduced oscillator strength of exciton.[31] In our work, the A$_{1g}$ mode is observed to soften and widen (402 cm$^{-1}$) at the monolayer MoS$_2$ Case 1 interface compared to that (406 cm$^{-1}$) of the Case 2 interface. We rule out the possibility of phase transition to metallic 1T MoS$_2$ since no notable peaks related to the 1T phase emerge in the range of 100-350 cm$^{-1}$ in addition to the above discussed reasons.[37,38] Considering this, we conclude that the monolayer MoS$_2$ interface with directly evaporated Au is degenerately doped, while MoS$_2$ interface with transferred Au is not.

To verify our Raman analysis, we performed first principles calculations using density functional theory (DFT) of the Raman modes for MoS$_2$ with and without an Au layer (figure 2g). The strength of the interaction between the MoS$_2$ and the metal surface was modelled by varying the distance between the surfaces and computing the interaction energy using van der Waals corrected DFT (See methods). Bader charge analysis suggests an increase in charge transfer with decreasing interlayer distance, and as a result, a modified Fermi level emerges for MoS$_2$. The electronic properties of the MoS$_2$ also change in the presence of a metal surface. The direct band gap (~1.66 eV semiconductor MoS$_2$, due to the presence of local strain and charge transfer from the surface, turns into an indirect band gap (~1.44 eV) semiconductor with significant reduction of the bandgap for an interlayer separation of 2.54 Å. This computational result is in agreement with the significantly suppressed photoluminescence spectrum of the Case 1 interface (Fig. 2b). In addition, the intimate contact with Au alters the Mo-S bond length depending upon which side of the MoS$_2$ surface is considered (the MoS$_2$ surface adjacent to, or on the opposite side of the Au layer). As a result, the Raman intensities of $E_{2g}^1$ and $A_{1g}$ are highly suppressed (figure 2g) near the Au which coincides with our far-field Raman spectrum (figure 2e). We note that Mo-S bonds are spatially varying along the layer thickness due to the presence of an Au layer on only one side, leading to slight shifts of $E_{2g}^1$ and $A_{1g}$ modes in MoS$_2$-Au contact (green and blue bars in figure 2g). In addition, charge density differences also proved strong electronic coupling between MoS$_2$ and Au by charge transfer (figure 2h). Significant charge transfer (~0.4e$^-$) occurs from Au to the vacant *d*-orbitals of the chalcogen atoms of MoS$_2$. The calculation result verifies that the charge transfer and the strain play a critical role in the quality, and hence electrical and optical properties of MoS$_2$-Au contact.

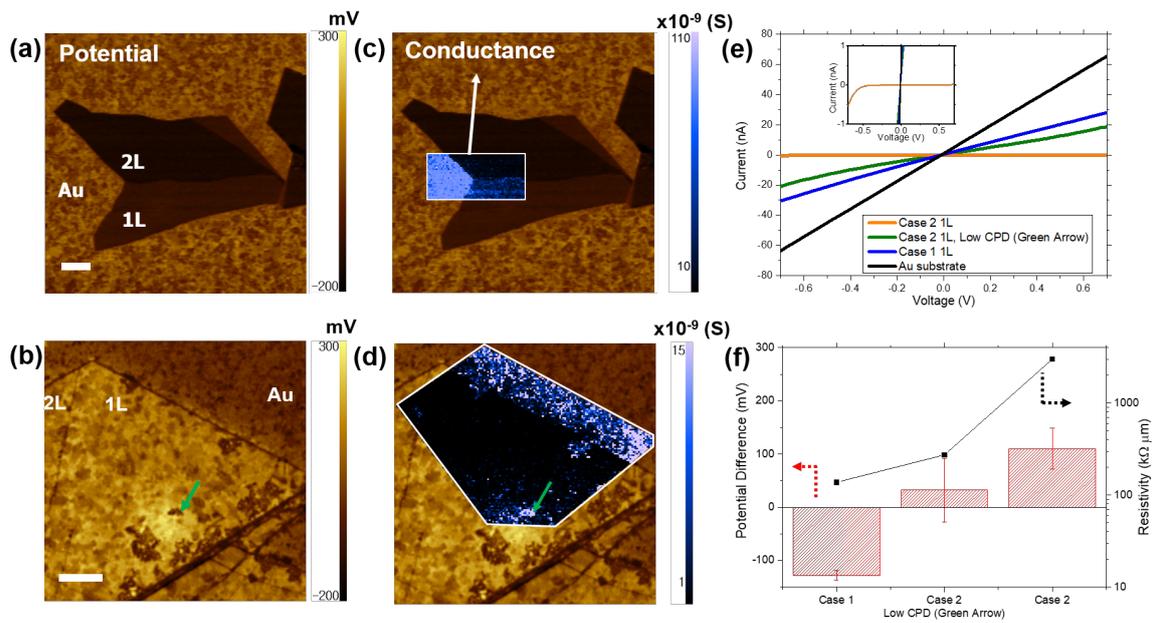

Figure 3. (a, b) Surface Potential maps, (c, d) Conductance mapping at -0.6 to -0.7 V of (a, c) for directly evaporated Au (Case 1) and (b, d) the transferred Au interface (Case 2). The green arrows in (b) and (d) indicate the identical position of Low potential region. (e) I-V Curves and (f) Potential difference between Au substrate and each interface plotted with corresponding resistivity. Scale bars at (a) and (b) indicate 1 μm.

Surface potential of the monolayer $MoS_2$ Case 1 and the Case 2 interface were investigated by KPFM. KPFM measures potential difference between a sharp tip in the proximity of a solid surface to obtain the work function difference for the solid. The potential at the 2D semiconductor-metal interface arises from both effects of charge transfer at the interface as well as the band structure of the semiconductor. Naturally, an interface dipole can be formed due to charges in the semiconductor and mirror image at metal. Moreover, since the $MoS_2$ thickness is much smaller than that of Thomas-Fermi screen length of 3 nm along the z-direction, the potential due to interface dipole is not screened out (see supplementary figure S3).[39] Therefore, we can directly measure the potential contributions by charge-transfer at the $MoS_2$-metal contact. We observe stark differences in our potential maps of $MoS_2$-Au interfaces for the two different types of contacts being investigated. The $MoS_2$ with directly evaporated

Au (Case 1) exhibits lower potential than the Au substrate, while the MoS$_2$ directly exfoliated on Au (Case 2) exhibits higher potential as compared to Au. The comparative potential difference (V$_{interface}$-V$_{Au\ substrate}$) is shown in figure 3f. The potential difference of the monolayer MoS$_2$ Case 1 interface compared to Au substrate is recorded to be -127 ± 8.73 mV and the corresponding work function of the monolayer MoS$_2$ was 5.29 eV, which is larger than Au substrate (figure 3a, f). As we discuss above, considerable strain of ~5.5% is imposed on MoS$_2$ underlying the directly evaporated Au. Prior theoretical estimates have suggested that strained MoS$_2$ with 4% biaxial (tensile) strain can eliminate any Schottky barriers since the conduction band minimum shifts below the Fermi level due to the increase of electron affinity.[23] Further, it was also reported that electron affinity of monolayer MoS$_2$ linearly changes as a function of strain to a value of 5.40 eV for ~5.5% strain. This should result in a significant shift of conduction band minimum far below the Fermi level considering the measured work function value of 5.29 eV. Since electron affinity is higher than the work function of Au, Schottky barrier height becomes negative according to the Schottky-Mott model. However, the Fermi level must be identical throughout the entire region, hence negative charges must accumulate on the MoS$_2$ side (see supplementary figure S3a). Assuming the type of charge induced by the MoS$_2$ dipole dominates the potential due to high order potential ~$z^{-3}$ where z is distance, we can extrapolate that negative dipole charges are the dominant contributors to the potential measurement on the Case 1 interface. Qualitatively, we can interpret the negative potential difference of -127 mV is attributed to the negative dipole at the MoS$_2$ side of the interface (figure 2h and figure S3a, b). In this respect, surface charge of the interface was readily predicted by KPFM. Therefore, we conclude that MoS$_2$ is degenerately *n*-doped mainly because of biaxial strain and a partial contribution of the negative interface dipole. *n*-type conduction at the MoS$_2$-Au interface is widely accepted and consistent with previous predictions by calculation.[40,41] The ohmic behavior at the Case 1 interface is consistent to previous predictions of ohmic behavior in clean electrical contacts.[42]

On the other hand, the directly exfoliated monolayer MoS$_2$ (Case 2) interface shows a positive potential difference, +110 ± 38.6 mV and a work function of 5.05 eV (figure 3b, f). As per the discussion above, ~0% strained MoS$_2$ should have electron affinity of 4.59 eV according to the literature.[23] Considering the Au workfunction of 5.16 eV (see method), a Schottky barrier is formed at this interface. Following the same reasoning presented above, positive charges collect on the MoS$_2$ side of the interface (see supplementary figure S3c, d). Hence it proves that the vdW maintains *n*-type semiconducting behavior but creates a resistive Schottky barrier type contact. Similar behavior was observed in the bilayer regions shown in figure 3(c, d) and are discussed in more detail later.

The stark contrast between the Case 1 and Case 2 interface is attributed to coupling between Au and S atoms. The overlap between *d*-orbitals of Au and MoS$_2$ is critical for contact.[40] This is because *d*-orbitals far overwhelm the *sp* orbitals in terms of density of states. Therefore, the distance between Au and S atoms can considerably change the conductivity or resistivity across the interface. Due to different coupling strength, not only the strain on the MoS$_2$ surface but also the dipole direction of the interface is changed. This interface effect emerges uniformly throughout the whole Case 1 interface, which results in a spatially uniform surface potential on the monolayer and the bilayer MoS$_2$. Interestingly, the monolayer MoS$_2$ region shows uniform potential throughout the flake, despite deviation of the potential value for Au itself (+28.7 ± 33.5 mV). Note that the Au substrate also shows spatial variation of its potential value due to the polycrystalline nature of evaporated Au.[43] The spatially uniform potential of the monolayer MoS$_2$ in Case 1 interface indicates that the electrical response of Au is mostly screened by the free carriers of MoS$_2$, regardless of pre-occupied surface states. On the other hand, the monolayer MoS$_2$ in Case 2 interface exhibits pronounced non-uniformity in surface potential distribution. This non-uniformity is attributed to spatially varying MoS$_2$-Au distance and correspondingly uneven dipole distribution. This agrees with the TEPL map (figure 2d) showing that junction quality between Au and sample is randomized due to the spatially varying MoS$_2$-Au distance. Interestingly, some regions of the Case 2 interface recorded dramatically low potential, +32.6 ± 59.6 mV (green arrow at figure 3b). We interpret this data as reduction of the Schottky barrier in these regions, resulting in higher conductivity.

Accordingly, the I-V response and conductance maps were collected for each interface under consideration here (figure 3e, f). I-V curves measured by C-AFM describe out-of-plane carrier transport. Therefore, this data can serve as a measure of direct carrier injection at 2D semiconductor-metal interface. Figure 3e represents spatially averaged I-V curves (6 x 6 points). The monolayer MoS$_2$ Case 1 interface showed a linear response of current to voltage, indicative of ohmic behavior. It is also consistent with potential map and reflectance measurements, further confirming our conclusion of degenerate doping due to direct Au evaporation. Similar to potential and TEPL maps, the Case 1 interface also shows good spatial uniformity of conductance. Prior research has provided some evidence using cross-sectional TEM imaging that direct metal evaporation leads to creation of point defects, metal diffusion, broken layers of the 2D chalcogenide, and glassy layers at the evaporated MoS$_2$-Au interface.[10] The authors attributed these defects to the bombardment by the high-energy Au atoms on the MoS$_2$ surface. We do not find any evidence of such defects or Au diffusion through the MoS$_2$ in the potential and conductance maps acquired by scanning probe. We have also performed cross-section TEM imaging to verify our claims (see supplementary figure S4). On the other hand, most of the monolayer MoS$_2$ Case 2 interface was highly resistive and showed rectifying behavior due to the presence of tunnel and Schottky barriers. This implies an obvious relationship between conductivity and potential at the interface (figure 3h). We extracted out-of-plane resistivity from I-V curves and contact area was obtained by using a Hertzian contact model (See Methods). Based on our calculation, the monolayer $MoS_2$ Case 1 interface reaches to 138 k$\Omega$ μm (potential difference of -127 mV) while the Case 2 interface recorded 3,009 k$\Omega$ μm (potential difference of +110 mV). It is worth nothing that there are localized low-potential regions on the Case 2 interface (green arrow) which exhibit a linear I-V relationship and comparable conductance to that of the Case 1 interface. The resistivity of these localized regions at the monolayer $MoS_2$ Case 2 interface was determined to be 274 k$\Omega$ μm (potential difference of +32.6 mV).

It is also worth noting that we do not find any evidence of tears, cracks, holes or other defects in topographic images of the sample at these locations of low-potential/high conductance. Therefore, we attribute these local "hotspots" of low-potential/high conductance to high-strain and/or interfacial charge that is randomly accumulated in the process of mechanical exfoliation.

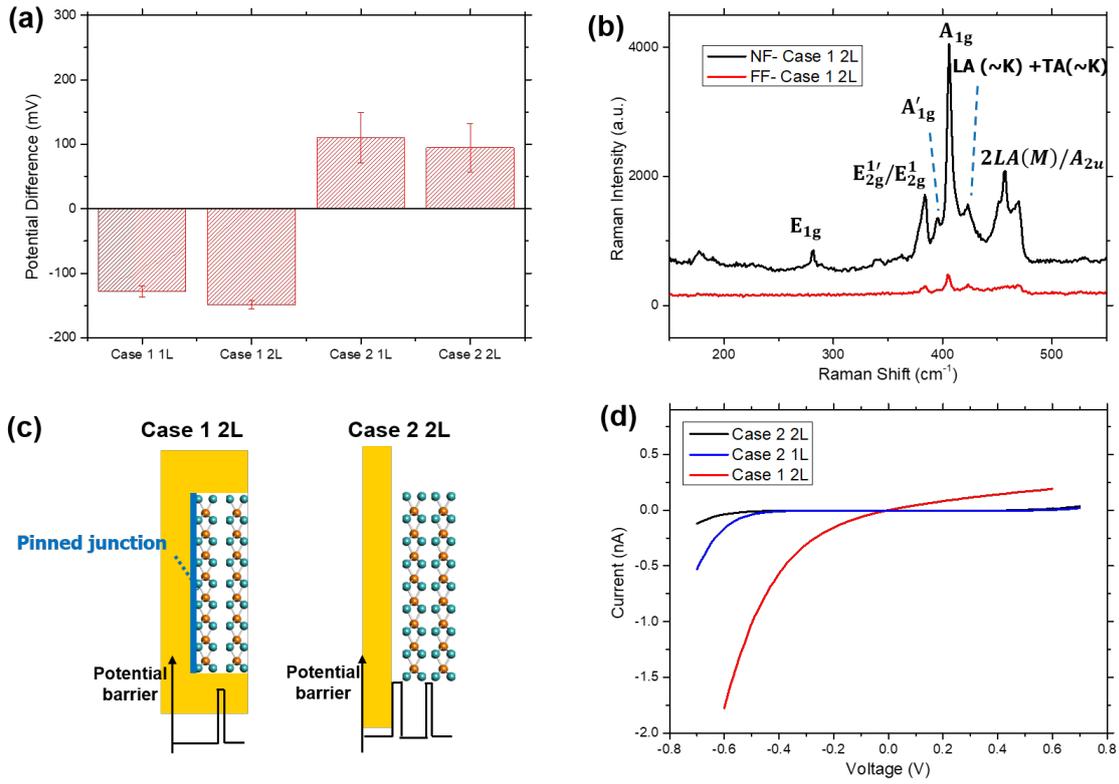

Figure 4. (a) Layer dependency of surface potential of $MoS_2$ w.r.t. Au substrate and (b) TERS spectrum of two types of contact with bilayer $MoS_2$ and Au. (c) Schematic diagram illustrating barrier potential and (d) I-V curves of the bilayer $MoS_2$ Case 1 and the Case 2 interfaces.

To further understand the impact of these two different types of contact formation processes between $MoS_2$ and Au we investigate the dependence of the number of $MoS_2$ layers on the Raman spectra and conductance of these contact interfaces. Bilayer $MoS_2$ presents an interesting case study to understand this further since a bilayer consists of two monolayers in two different surrounding environments; one layer is in direct contact with Au while the other layer is in contact with the first layer and faces the surface. Therefore, we expect significant changes in electrical properties between the two types of contact samples considered above. Figure 4a represents the potential difference of each contact interfaces comparing the monolayer and the bilayer MoS$_2$. Overall, the bilayer shows a small negative shift in potential regardless of interface type. We interpret that this shift is caused by layer dependent band offset and changes in the band gap. Considering that the doping concentration remains the same throughout the flake, the Fermi level in bilayer MoS$_2$ must shift in response to the reduction in bandgap. Per literature precedent, the bilayer conduction band drops by 0.1 eV and the valence band moves up by 0.2 eV as compared to the monolayer MoS$_2$.[44] Therefore, electron affinity and work function of bilayer MoS$_2$ increases and the corresponding surface potential drops. Quantitatively, potential shifts of -20.8 mV and -15.6 mV were observed for the bilayer MoS$_2$ Case 1 interface and the vdW interface, respectively. Moreover, the sign of the potential difference in bilayer MoS$_2$ samples (-148 ± 6.93 mV for Case 1 vs +94.6 ± 37.0 mV for case 2) was same as the monolayers for the same type of contact.

Once again TERS was used to investigate the degree of strain in bilayer MoS$_2$ under directly evaporated Au. Similar to monolayer MoS$_2$, strain-induced peak separation was analyzed and a compressive strain of 1% was deduced based on prior reports of strain vs Raman peak shift relations (figure 4b, supplementary figure S2a and table S1).[34] Compared to the 5.5% strain estimated in monolayer MoS$_2$ buried under directly evaporated Au, the strain in the bilayer is considerably suppressed. This can be attributed to the weak interlayer coupling in between MoS$_2$ layers of a bilayer sample which helps relieve some strain.

Out-of-plane conductance measurements of bilayer MoS$_2$ (figure 4 d) samples on Au add further insights and evidence to the above claims. On the other hand, the bilayer MoS$_2$ Case 1 interface only includes interlayer spacing (figure 4c, left) with a pinned junction due to strong coupling between Au and S. The bilayer MoS$_2$ Case 2 interface shows rectifying behavior from two layers of potential barriers in series (figure 4c, right). Since both junctions act as tunneling barriers for the bilayer MoS$_2$ Case 2 interface, current is reduced compared to the monolayer

MoS$_2$ Case 2 interface. Strikingly, the bilayer MoS$_2$ Case 1 interface shows a remarkable increase in current for the same driving voltage and a linear ohmic contact like behavior. This can be understood by considering a weakly coupled semiconductor layer to a pinned, degenerately doped semiconductor with a high work function (figure 4c, right). This type of contact may be comprised of relatively narrow Schottky and tunnel barriers that do not produce much rectifying behavior, resulting in efficient charge injection and low-contact resistance.[41,45]

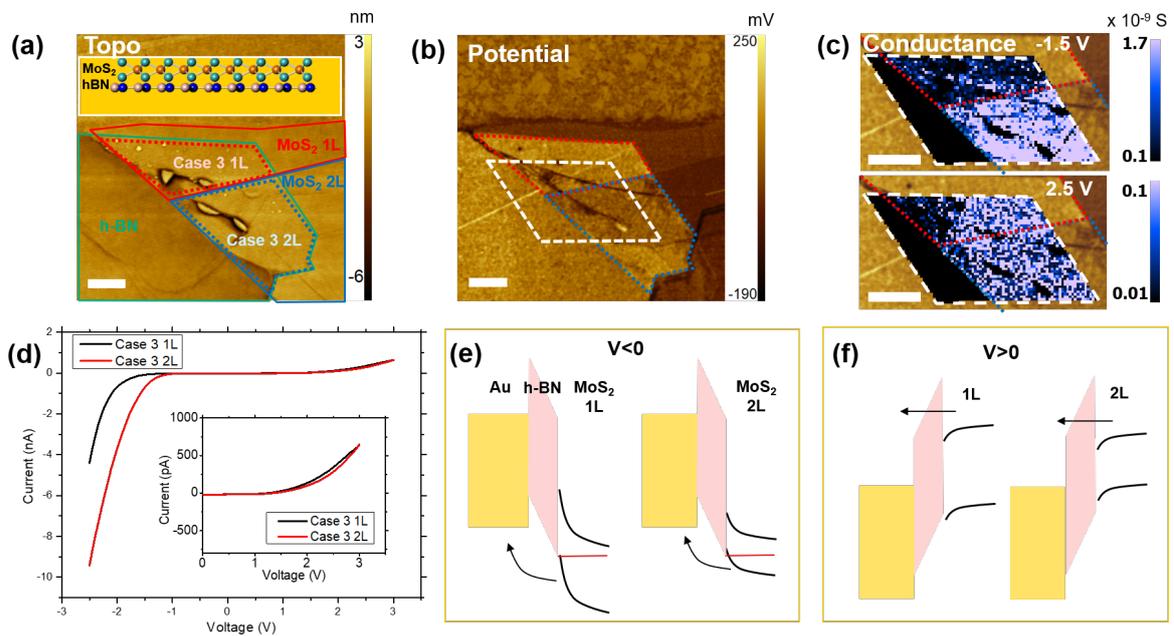

Figure 5. (a) Topography of MoS$_2$/h-BN/Au interface (Case 3). Inset in (a) is schematic representation of the Case 3 interface. Red dotted lines correspond to the monolayer MoS$_2$ Case 3 while blue dotted lines correspond to bilayer MoS$_2$ Case 3 interface. (b) Potential (c) Conductance map of the Case 3 interface with (top) -1.5 V and (bottom) +2.5 V at area covered by white dashed line in (b). Background is potential map. (d) I-V curves of mono and bilayer Case 3 interface. (e, f) Band diagram of the Case 3 interface at (e) V < 0 and (f) V > 0. Scale bar at all figures indicates 1 μm. Solid lines in (a) indicate monolayer MoS$_2$ (red), bilayer MoS$_2$ (blue) and h-BN (green).

To further understand the nature of buried interface between directly evaporated Au and MoS$_2$

we need to develop a way to tailor/modify this interface to eliminate this degenerate doping of monolayer in immediate proximity to MoS$_2$ and de-pin the Fermi level. As discussed above, Fermi level pinning is one of the critical challenges to address in 2D semiconductor/3D metal contacts.[7] A strong electronic bond between metal and semiconductor materials is necessary for effective charge transfer doping and reduction of contact resistance, yet, at the same time, metal-semiconductor bonding results in Fermi level pinning. Many experimental and computational approaches to eliminate Fermi level pinning have been proposed including placing interlayers such as graphene and hexagonal boron nitride (*h*-BN) between metal and TMDCs.[46–48] To investigate this de-pinning effect and corresponding influence on cross plane-carrier transport, we fabricated a MoS$_2$/*h*-BN/Au buried interface (Case 3) by taking advantage of Au-assisted dry-transfer technique (figure 5a, Fabrication details in Method). Briefly, the presence of trilayer *h*-BN on MoS$_2$ prevents Fermi level pinning by avoiding direct contact with Au during evaporation (inset at figure 5a and figure S5a-d). The *h*-BN interlayer not only helps achieve a de-pinned contact interface, but also helps in strain relaxation as evident from the Raman spectrum of the monolayer MoS$_2$ Case 3 interface (see supplementary figure S5e). This is expected since MoS$_2$ is not directly in contact with Au but rather with *h*-BN, which has similar lattice constant and weak van der Waals interaction.[49] In addition, there is no significant charge transfer between the metal and MoS$_2$ since they are physically separated by ~1 nm of a wide-gap insulator. Figure 5b represents the surface potential map of this buried interface heterostructure, comprising regions of both monolayer (red) and bilayer (blue) MoS$_2$ spaced by ~1 nm *h*-BN from the Au contact. Both regions showed obviously higher potential than the MoS$_2$ region without *h*-BN. The potential value of the monolayer (red dotted area) and bilayer (blue dotted area) MoS$_2$ with the Case 3 interface recorded marginally different values compared to Case 1 (+50.6 mV for the monolayer MoS$_2$ Case 3 and -5.57 mV for the bilayer

MoS$_2$ Case 3 interface). Corresponding work functions were 5.11 eV for monolayer MoS$_2$ Case 3 and 5.16 eV for the bilayer Case 3 interface. The values were between those measured for Case 1 and Case 2 MoS$_2$ monolayers. This verifies that the de-pinned MoS$_2$ work function is very different from that of the Case 1 interface. We want to note that there is an unexpected local strain during *h*-BN transfer leading to a wrinkle (stripe in potential map) across the *h*-BN and Case 3 interface. Likewise, trapped air/hydrocarbon bubbles may also cause potential turbulence. TEPL further reveals that the monolayer MoS$_2$ Case 3 interface is electronically isolated from the Au and emits high PL while the PL from all other regions are quenched, including from bilayer MoS$_2$ (See supplementary figure S5f). Considering the above, it is clear that the electrical properties of the Case 3 interface are fundamentally different because effects from neither Fermi level pinning nor strain are manifest throughout the MoS$_2$ flake. To understand better, we performed conductive AFM measurement by sweeping the applied voltage to the sample from -2.5 to 3 V (figure 5c-f). Overall, the Case 3 interface showed lower conductance than the directly contacted region (see supplementary figure S6b) as expected. Interestingly however, the conductance was strongly dependent on MoS$_2$ layer number in the negative bias (figure 5c top, d). Conduction through bilayer MoS$_2$ starts to increase at lower overpotential as compared to monolayer MoS$_2$. This is attributed to band alignments of these metal insulator semiconductor tunnel junctions where charge transport is over the barrier and hence, band alignment of the semiconductor with the barrier insulator is critical. Compared to the monolayer MoS$_2$ bandgap, the bilayer MoS$_2$ band gap is lowered by a conduction band decrease of ~0.1 eV, and valence band increase of ~0.2 eV.[44] Moreover, it has been reported earlier that *h*-BN/MoS$_2$ alignment imparts large barrier height for electrons but is negligible for hole transport.[50] Therefore, the current under negative bias results not from tunneling but rather from hole diffusion through the band inversion (figure 5e). Note that after band inversion by sufficient voltage (-2.5 V), both regions reached similar conductance (figure S6a). On the positive voltage side, similar conductance is observed throughout, regardless of MoS$_2$ thickness (figure 5c bottom, d). Further, the current value was smaller compared to hole conduction. This can be attributed to the through-barrier (trilayer *h*-BN) tunneling mechanism for charge transport (figure 5f) in a metal-insulator-semiconductor junction where the band gap of the semiconductor makes no significant difference. The tunneling behavior was further verified by introducing thicker *h*-BN barrier layers between MoS$_2$ and Au (see supplementary figure S7), which shows clear dependency on *h*-BN barrier thickness, independent of the number of MoS$_2$ layers, further verifying our claims.

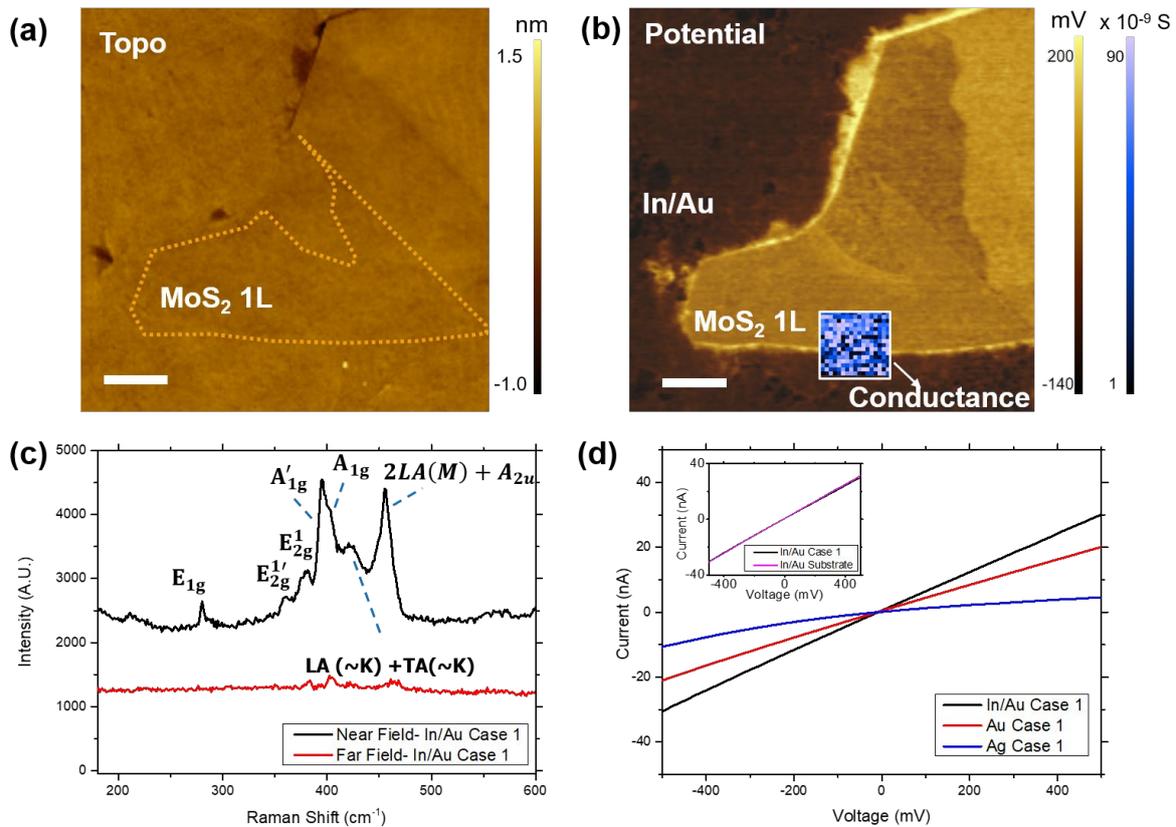

Figure 6. (a) Topography (b) potential and conductance map (c) Raman spectrum and (d) I-V curves of monolayer MoS$_2$ - In/Au Case 1 interface (The region surrounded by orange dotted line at figure 6a). Inset is I-V curve

comparison of MoS$_2$- In/Au Case 1 interface and In/Au substrate. Spatial conductance map was extracted using a voltage range of -0.4 to -0.5 V. Scale bar at (a) indicates 1 μm.

To further enrich our understanding of the MoS$_2$-metal interface, we have generalized our scheme to other metal contacts aside from Au using the same transfer technique of template stripping. This technique is applicable to other metals such as In and Ag (figure 6a and S8a) which have a reduced tendency to form oxides in contrast to other common contact metals. Further both In and Ag have been evaluated in the past as contact metals for MoS$_2$. In particular, a recent demonstration of In/Au alloy or In contacts with monolayer MoS$_2$ revealing exceptionally low contact resistance,[11,51] motivates investigation of this contact type. As indicated in figure 6a, the In/Au Case 1 interface was also flipped over and shows a flat surface. Details of evaporation and stripping of In/Au layer from the SiO$_2$/Si wafer are provided in methods and supplementary information. A surface potential map of the In/Au alloy contact indicates that electronically uniform contact was formed throughout the monolayer region, (figure 6b). The high potential of +114 mV at the buried interface region was observed. It is attributed to the low work function of In (4.12 eV) immediately beneath the MoS$_2$ layer.[52] Interestingly, the potential of the In/Au substrate was lower than the In- monolayer MoS$_2$ Case 1 interface and similar to that of Au substrate (~0V). Therefore, the potential map contrast (Figure 6b) is opposite to the contrast of the pure Au contact buried interface (Figure 3a). This indicates that Au atoms are the dominant surface facing species at the In/Au substrate, while the buried interface with MoS$_2$ maintains In as the primary contact element facing MoS$_2$. We speculate that the diffusion rate of Au on MoS$_2$ during annealing is different from SiO$_2$ for such a condition to arise. Near-field and far-field Raman spectra were collected via TERS to investigate strain and charge transfer effects at this In-Au alloy interface with MoS$_2$ (figure 6c).

Similar to the buried interface with pure Au, the far field signal was suppressed, while the near-field signal was enhanced by the Purcell effect of the In/Au plasmonic nanocavity with the Au scanning probe tip. Strain analysis of the Raman peaks suggests a ~5% (tensile) strain at the In/Au interface with $MoS_2$, comparable to the Au buried interface. Finally, local conductance measurements reveal a linear I-V relationship which shows ohmic contact (figure 6d). Remarkably, the z-directional conductance was much higher than the Au buried interface and almost identical to that of In/Au substrate (inset of figure 6d), in agreement with prior reports of low contact resistance in In/Au contacted $MoS_2$ FETs. Accordingly, the extracted conductivity reached 63 k$\Omega$ μm for 1L.

Likewise, the Ag buried interface was also investigated and unlike Au and In/Au, Ag contacts show much higher contact resistance (see supplementary figure S8) despite the lower work function as compared to Au. This suggests that contact resistance is largely influenced by Fermi level pinning rather than metal work function. Fermi level pinning can be addressed by careful control over the interface structure and evaporation conditions. The Ag contacts applied via thermal evaporation unlike the e-beam evaporated Au contacts likely created additional damage to the $MoS_2$ layer during metal cooling. For comparison, thermally evaporated Au contacts were characterized and demonstrated similar I-V curves and resistivity (see supplementary figure S8e), suggesting the contact application method strongly dictates the performance of the contact.. This is also evident in the surface potential maps of thermally evaporated Au and Ag where, huge work function variation was observed which was noticeably absent in e-beam evaporated Au and In/Au (figure S8b, d and figure 3a, 6b).

Discussion and Conclusions

We report a thorough and comprehensive study on the buried interface of a prototypical 2D semiconductor (MoS$_2$) with 3D metals (Au, In/Au and Ag). In particular, we investigate comparative differences between Au/MoS$_2$ contact formation via two different means: 1. Direct Au evaporation on the semiconductor and 2. Direct exfoliation of the semiconductor on freshly stripped Au. Using a combination of scanning probe techniques and far-field as well as near-field optical techniques we have directly probed the optical and electrical properties of 2D semiconductor-metal interface and inferred the conclusions to understand nature of contact resistance at this buried interface. Based on our observations and interpretation, direct metal evaporation is undoubtedly better in terms of providing an electrically uniform, homogeneous contact as compared to direct exfoliated semiconductor on a metal surface. Direct metal evaporation results in large strain in the buried monolayer and bilayer MoS$_2$ which results in significant changes to bandgap and corresponding electrical contact properties. Fermi level pinning is the dominant effect in direct metal evaporation which also affects the surface potential of the second layer bilayer MoS$_2$. In contrast, electrical contact in directly exfoliated samples is spatially inhomogeneous and electrically poor (low conductance) overall with the exception of a few hotspots of low surface potential/high conduction. Presence of an isolating interlayer such as h-BN between the MoS$_2$ and Au relieves strain and de-pins the Fermi level on MoS$_2$. Use of soft metal contacts such as In/Au reveals that conductance through the buried interface can be significantly improved. Likewise, use of thermally evaporated Au and Ag contacts suggests that conductance through the contact increases suggesting the cooling and growth of metal after landing on semiconductor has a defining role to play in determining the contact resistance. It is clear from our observations that while direct metal evaporation is the favored method for making contacts with low interface resistance, direct metal evaporation also results in Fermi level pinning at the 2D semiconductor/3D metal interface. To overcome this drawback one either needs an intermediate de-pinning barrier layer such as *h*-BN or a soft, non-reactive low-melting point metal such as In. In addition, the nature of metal evaporation and post-condensation cooling on the semiconductor also has an important role to play in determining contact resistance. Our results suggest that e-beam evaporation in combination with use of ultrathin barrier layers and low-melting point metals hold the most desirable recipe towards forming de-pinned, low-resistance contacts of bulk metals to 2D semiconductors.

In conclusion, we have performed a detailed scanning probe-based study of the buried 2D semiconductor/3D metal interface and identified various attributes of the interface that contribute to the quality of electronic contact at the interface. We have also identified various criteria related to evaporation conditions and materials needed to form low-resistance metal contacts to 2D semiconductors.

## Methods

*Fabrication of MoS$_2$-Metal (Au & Ag) Direct Evaporation (Case 1) Interface*

The sample fabrications process is identical to the process described in a previous paper.[22] Briefly, mono- and bilayer MoS$_2$ were mechanically exfoliated on Si/SiO$_2$ substrate. 100 nm of Au layer was deposited on as-exfoliated MoS$_2$ by either e-beam evaporator (Kurt J. Lesker PVD 75 PRO-Line E-Beam Evaporator) or thermal evaporator (Kurt J. Lesker Nano 36 Thermal Evaporator). For Ag, thermal evaporator was used (Kurt J. Lesker PVD 75–e-Beam/Thermal Evaporator). Deposition rate was set to 0.2 nm/s under low pressure ($5 \times 10^{-7}$ torr). Epoxy was applied on Au layer and another Si substrate was placed as a transfer substrate. After heating up to 100 °C for 2 hrs, the transfer substrate was gently peeled off, resulting in buried MoS$_2$ on Au or Ag.

*Fabrication of MoS$_2$ -In/Au Case 1 interface*

6 nm of In and 80 nm of Au layers were evaporated on the as-prepared MoS$_2$ on Si/SiO$_2$ substrate by e-beam evaporator (Denton Explorer 18). Deposition rate was set to 0.2 nm/s with the substrate holder water cooled at 15 °C. The sample was annealed at 300 °C, N$_2$ ambient condition for 3 hrs. Right after that, it was immersed in acetone and sonicated for 5 s. Using blade, edge of the transfer wafer was scratched so that it was able to peel off.

*Fabrication of MoS$_2$/h-BN/Au Case 3 interface*

First, few layered MoS$_2$ is mechanically exfoliated on Si/SiO$_2$ substrate. Few layered *h*-BN was transferred on top of the MoS$_2$. It is followed by Au evaporation. After that, identical process to Case 1 was conducted.

*Far-field reflectance measurement*

Reflectance from the sample was measured by illuminating a white light source on the sample and passing the reflected light through 600 grooves/mm grating before collecting them in a LabRam-EVO Raman spectrometer. Reflectance acquired from the polished silver mirror with the same acquisition setting, was used to normalize the reflectance from the sample to avoid any absorption from the gold substrate.

*Kelvin probe force microscopy (KPFM) & Conductive-AFM (C-AFM) measurement*

OmegaScope-R (AIST-NT) setup was used for KPFM and C-AFM measurement. For KPFM measurement, Au tip was biased by 3 V and connected to lock-in amplifier while sample was grounded. Tip was calibrated with freshly cleaved HOPG. Au substrate displayed work function of 5.16 eV in both samples, indicating Au layer was textured mainly along (100) direction.[43] For C-AFM measurement, tip was grounded while sample is connected to 10 MΩ series resistance and biased. I-V curves were collected and averaged from 6x6 points.

*Tip-enhanced Raman/Photoluminescence (TERS/TEPL) measurement*

Identical AFM setup as described above was coupled to a LabRam-EVO Raman spectrometer (Horiba Scientific), with a laser excitation, at 633 nm and with power of ~460 µW. Au coated OMNI-TERS probes (APP Nano) were used for all measurements.

*Resistivity Calculation*

To obtain contact area between tip and sample, Herztian contact model was used as literature described.[53]

$$d^{\frac{3}{2}} = \frac{3}{4} F \frac{1}{E^*} r^{-\frac{1}{2}}$$

(*d* is dented thickness by applied force, *F* is applied force, *E\** is reduced Young's modulus, *r* is tip radius)

$$\frac{1}{E^*} = \frac{1-v_1^2}{E_1} + \frac{1-v_2^2}{E_2}$$

($E$ is Young's modulus and $v$ is poisson ratio)

Young's modulus (Poisson ratio) of 1L MoS$_2$ and Au are 270 GPa (0.27) and 106 GPa (0.3), respectively.[54,55] Based on the values, reduced Young's modulus of MoS$_2$/metal samples were calculated. Plugging applied force in the equation during contact mode, $d$ is obtained 5.41 x 10$^{-11}$ m. Contact area ($S$) was calculated by $S = 2\pi r d$.

Van der Waals corrected Density Functional Theory (DFT) Calculation

We have performed all of the van der Waals corrected Density Functional Theory (DFT) calculations using the Vienna Ab initio Simulation Package (VASP)[56,57] with projected-augmented wave (PAW)[58] pseudopotentials and the Perdew–Burke–Ernzerhof (PBE) exchange–correlation functional.[59] We have used 5X5X1 k-points for the optimization processes and 9X9X1 k-points for n-SCF calculations for the MoS$_2$-Au (111) composite as well as parent systems. A 25 Å vacuum is employed in the z-direction to avoid all unwanted interactions between the layers. A plane-wave basis set energy cutoff of 400 eV was used for the calculations and the systems were optimized until forces on each atom were converged to less than 0.01 eV Å$^{-1}$. We have used a Γ-centered k-point mesh to calculated static phonon modes and Raman intensity. The dielectric constant and the optical absorption coefficient values of the composites using are also calculated using VASP. The calculations use the direct inter-band dipolar transitions.

# AUTHOR INFORMATION

**Corresponding Authors**

*dmj@seas.upenn.edu

**Author Contributions**

All authors contributed to this study. K. J. conducted scanning probe technique measurement and wrote the manuscript. K. J. and J. O. fabricated the samples. P. K. contributed to h-BN transfer on $MoS_2$ and electron microscopy. S. B. A. contributed to reflectance measurement and K.K. contributed to TEM sample prep. J.M., M. M., C. M. and N. G. assisted with sample preparation including In and Au evaporation. A.B. and V. S. contributed to DFT calculations. E.A.S. supervised the electron microscopy. D. J. supervised the entire research and was the thought leader on the study along with K. J. All authors contributed to revising the manuscript and interpretation of data/results.

**Funding Sources**

D.J. and K.J. acknowledge primary support for this work by the Air Force Office of Scientific Research (AFOSR) FA9550-21-1-0035 and U.S. Army Research Office under contract number W911NF-19-1-0109. D.J. and J.M. also acknowledge partial support from and FA2386-20-1-4074. D.J., E.A.S. and P. K. acknowledge support from National Science Foundation (DMR-1905853) and support from University of Pennsylvania Materials Research Science and Engineering Center (MRSEC) (DMR-1720530) in addition to usage of MRSEC supported facilities. The near-field work and metal evaporation was carried out at the Singh Center for Nanotechnology at the University of Pennsylvania which is supported by the National Science Foundation (NSF) National Nanotechnology Coordinated Infrastructure Program grant NNCI-1542153. J. O. was supported by the NSF-REU program hosted by the NNCI at Penn. K.K. acknowledges support for TEM sample preparation at the Center for Functional Nanomaterials, Brookhaven National Laboratory, which is a U.S. DOE Office of Science Facility, at Brookhaven National Laboratory under Contract No. DE-SC0012704. S.B.A acknowledges support from Swiss National Science Foundation Early Postdoc Mobility Program (187977). N.G. acknowledges support of Air Force Office of Scientific Research grant FA9550-19RYCOR050.

## Supplementary Information

Supplementary information contains actual value of reflectance, Deconvolution of Raman spectrum and Raman mode table, Schematic representation of KPFM vs band alignment, Cross-section HAADF STEM and EDS images, Characterization of $MoS_2$/hBN-Au Case 3 interface and Conductance, potential map of Ag Case 1 interface and Au Case 1 interface made from thermal evaporator.

Supplementary Information

# Direct Opto-Electronic Imaging of 2D Semiconductor- 3D Metal Buried Interfaces

Kiyoung Jo[1], Pawan Kumar[1,2], Joseph Orr[3], Surendra B. Anantharaman[1], Jinshui Miao[1], Michael Motala[4,7], Arkamita Bandyopadhyay[2], Kim Kisslinger[5], Christopher Muratore[6], Vivek B. Shenoy[2], Eric Stach[2], Nicholas Glavin[4], Deep Jariwala[1]*

[1]Electrical and Systems Engineering, University of Pennsylvania, Philadelphia, PA, 19104, USA

[2]Materials Science and Engineering, University of Pennsylvania, Philadelphia, PA, 19104, USA

[3]Electrical and Computer Engineering, Villanova University, Villanova, PA, 19085, USA

[4]Air Force Research Laboratory, Materials and Manufacturing Directorate, Wright-Patterson AFB, Columbus, OH, 45433, USA

[5]Brookhaven National Laboratory, Upton, NY, 11973, USA

[6]University of Dayton, Dayton, OH, 45469, USA

[7]UES Inc., Beavercreek, OH, 45432, USA

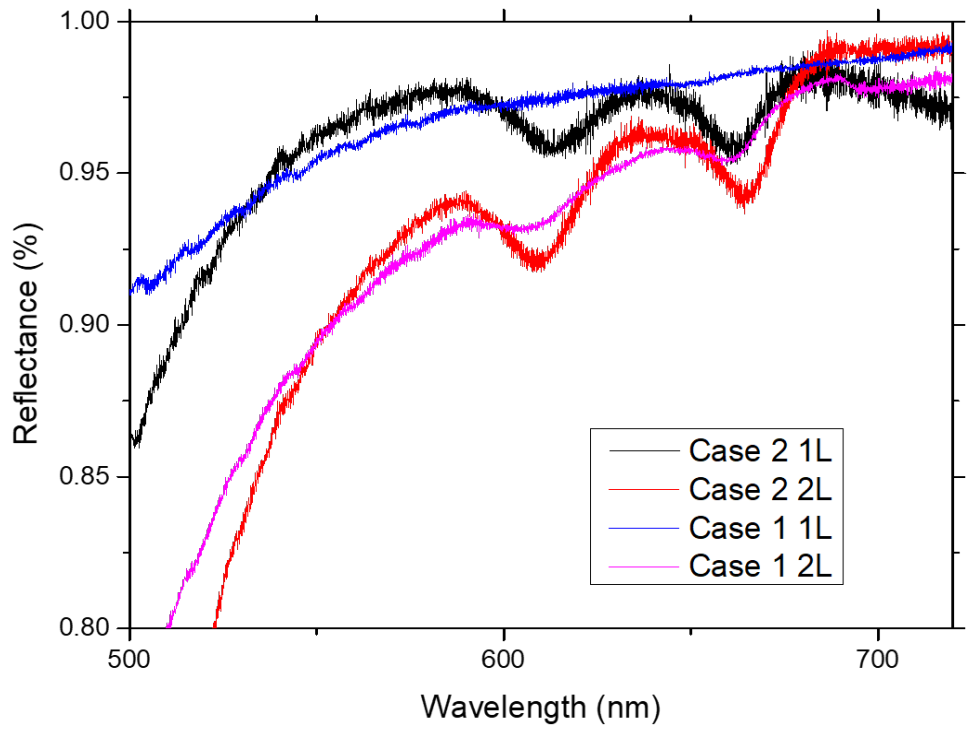

Figure S1. Reflectance spectrum with actual reflectance % value of MoS$_2$-Au Case 1 interface and the Case 2 interface.

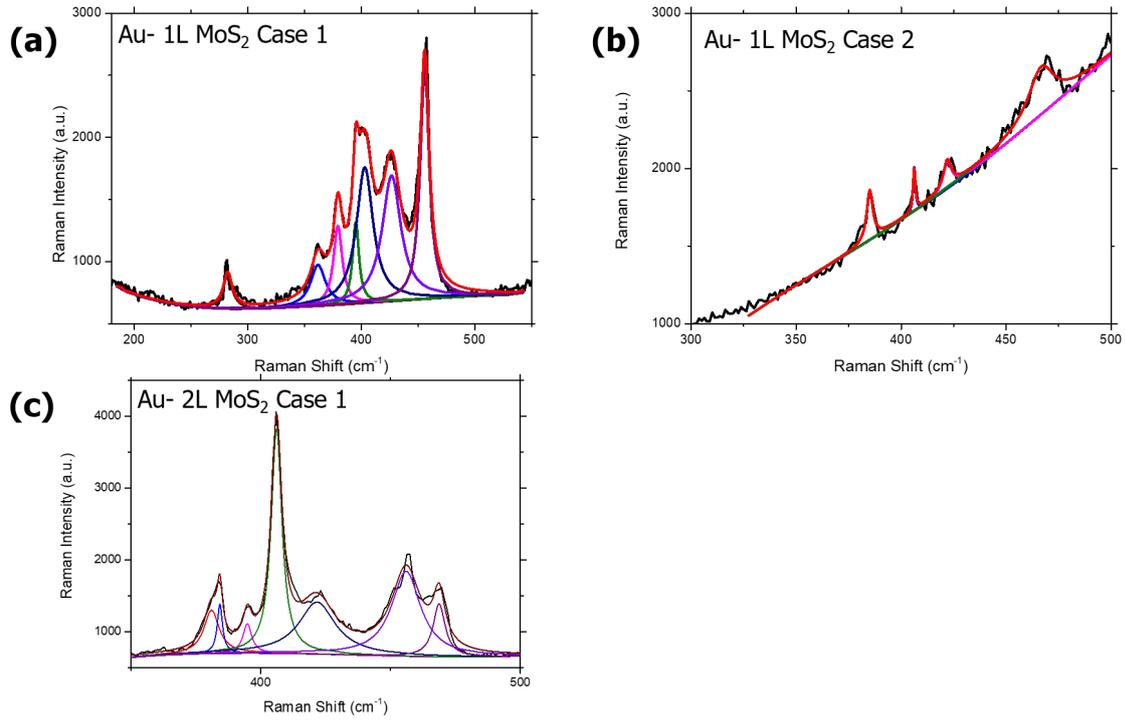

Figure S2. Deconvoluted Raman spectrum of (a) Au- 1L MoS$_2$ Case 1 (b) Au- 1L MoS$_2$ Case 2 and (c) Au- 2L MoS$_2$ Case 1 interface.

| Raman Shift (cm$^{-1}$) | Au Case 1 1L (FWHM) | Au Case 2 1L (FWHM) | Au Case 1 2L (FWHM) |
|---|---|---|---|
| $E_{1g}$ | 282.3 (9.613) | N.A. | 282.3 |
| $E_{2g}^{1\prime}$ | 361.8 (15.80) | N.A. | 381.1 (7.000) |
| $E_{2g}^{1}$ | 379.3 (9.394) | 385.1 (3.500) | 384.3 (2.443) |
| $A_{1g}^{\prime}$ | 395.1 (6.775) | N.A. | 394.9 (4.104) |
| $A_{1g}$ | 402.0 (17.79) | 406.3 (1.294) | 406.14 (5.012) |
| $LA(K) + TA(K)$ | 426.7 (20.01) | 422.0 (4.677) | 421.7 (17) |
| $A_{1g} + E_{2g}^{2}$ | 455.8 (9.464) | 466.7 (15.00) | 456.0 (13.59) |

Table S1. Raman mode of the buried interface and the vdW interface collected by tip-enhanced Raman spectroscopy.

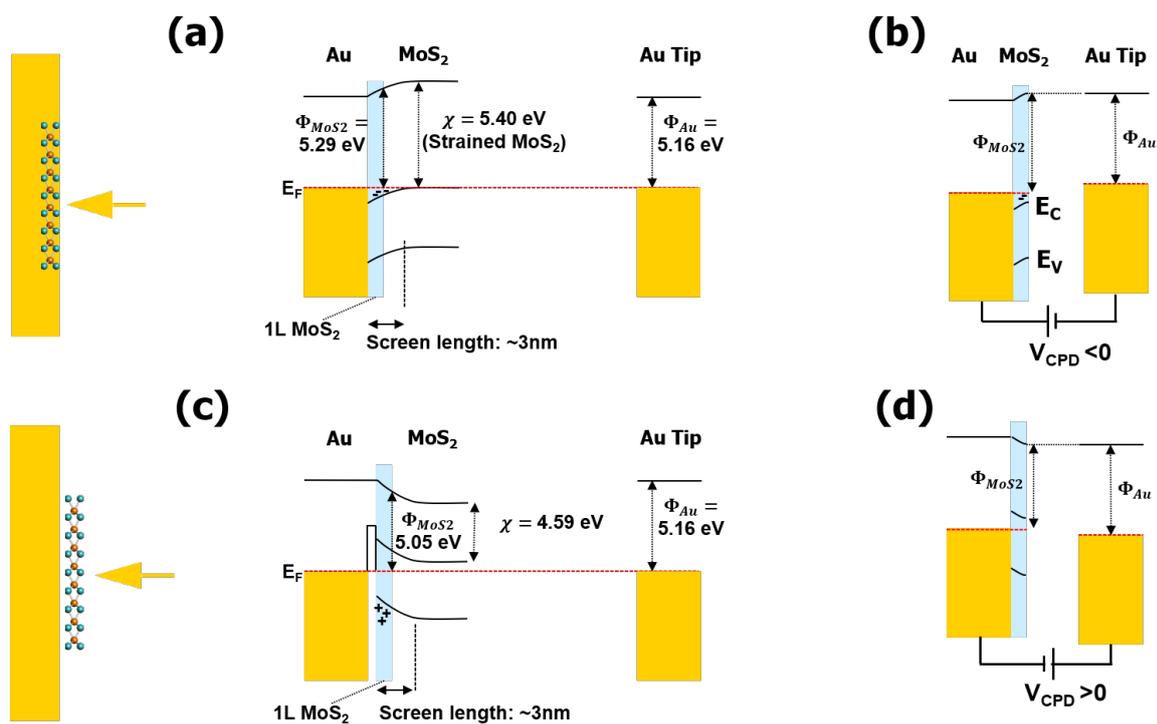

Figure S3. Band alignment of the buried interface and Au tip (a) without contact and (b) with KPFM feedback. (c) Band alignment of the vdW interface and Au tip (a) without contact and (b) with KPFM feedback.

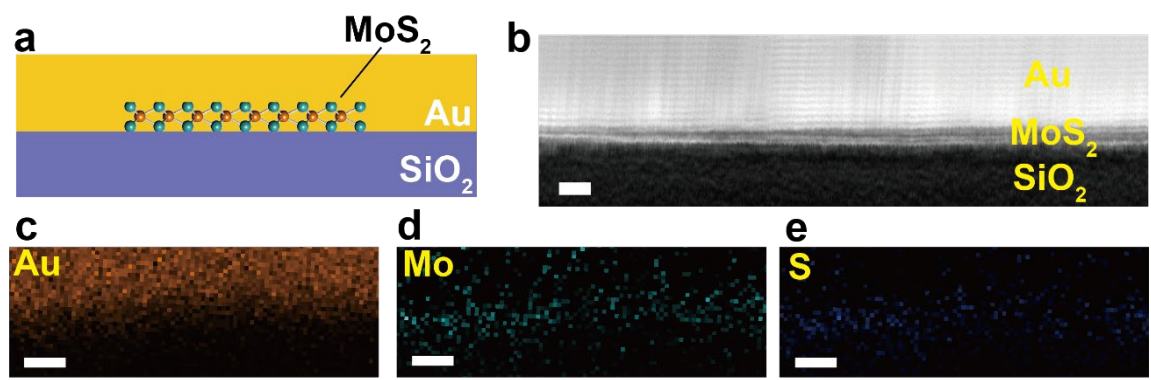

Figure S4. (a) Schematic representation of the MoS$_2$-Au buried interface before flipping. (b) Cross-section high-angle annular dark field (HAADF) scanning transmission electron microscopy (STEM) image of the buried interface and (c,d,e) Energy dispersive X-ray spectroscopy map of (c)Au, (d)Mo and (e) S. Scale bars indicate 1 nm.

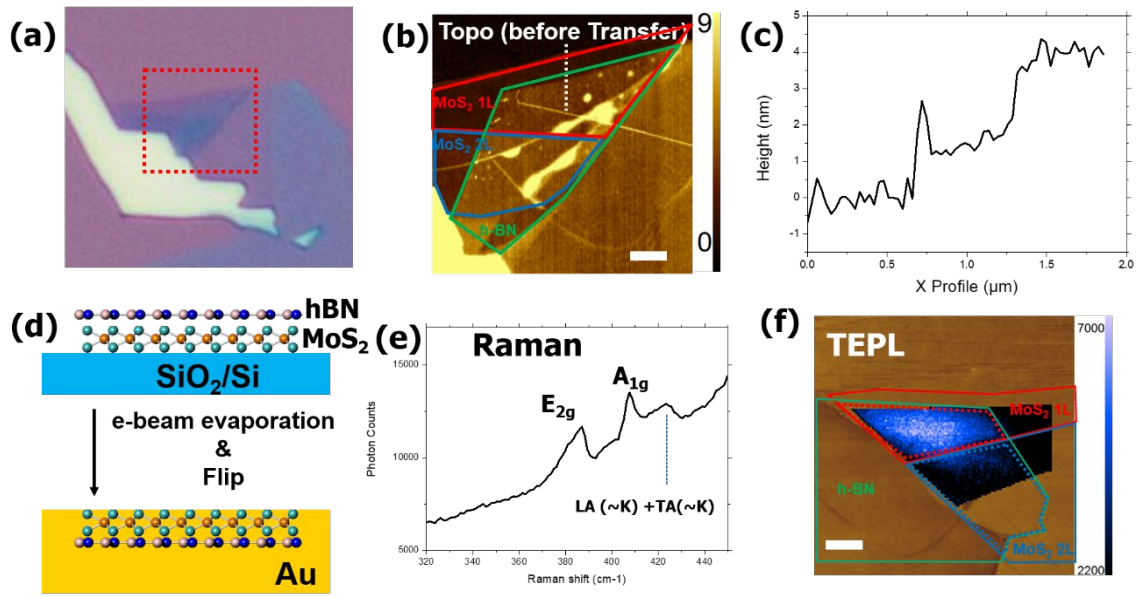

Figure S5. (a) Optical image and (b) Topography map (c) and corresponding height profile before Au-assisted transfer of hBN/MoS$_2$ heterostructure. (d) Representative illustration of hBN/MoS$_2$ structure and MoS$_2$/hBN-Au buried interface. (e) Raman spectrum and (f) TEPL map of the buried interface. Scale bar indicates 1 μm.

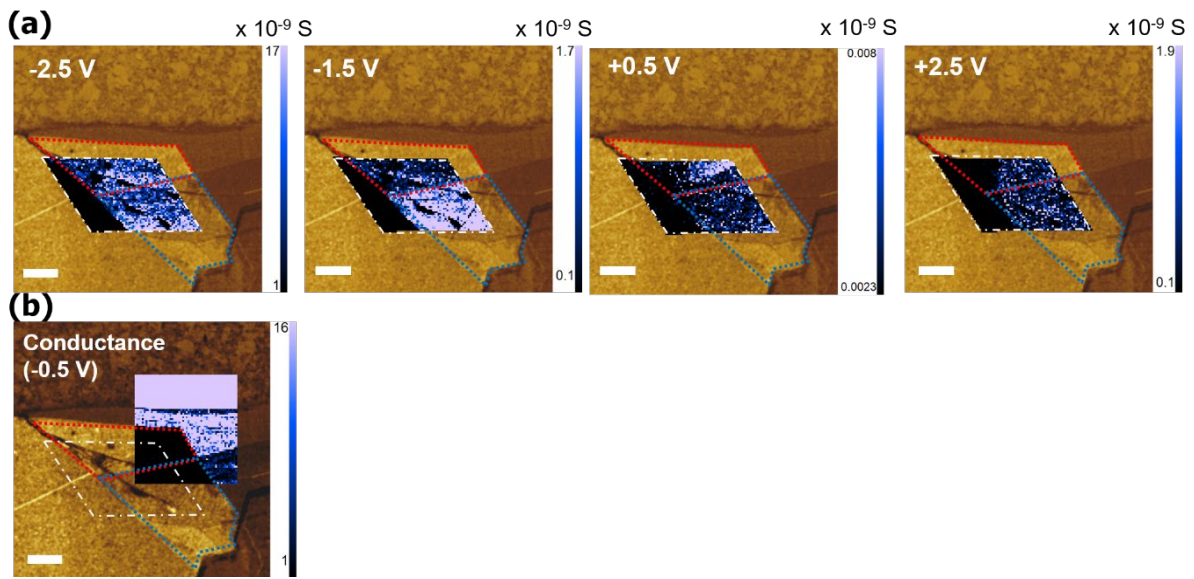

Figure S6. (a) Conductive map of MoS$_2$/hBN/Au buried interface (Case 3) from -2.5 V to +2.5 V. (b) Conductance comparison covering Au, MoS$_2$/Au Case 1 interface and MoS$_2$/hBN/Au Case 3 interface. Scale bar indicates 1 μm. 1L MoS$_2$ Case 3 is marked by red dotted line and 2L MoS$_2$ Case 3 is by blue dotted line.

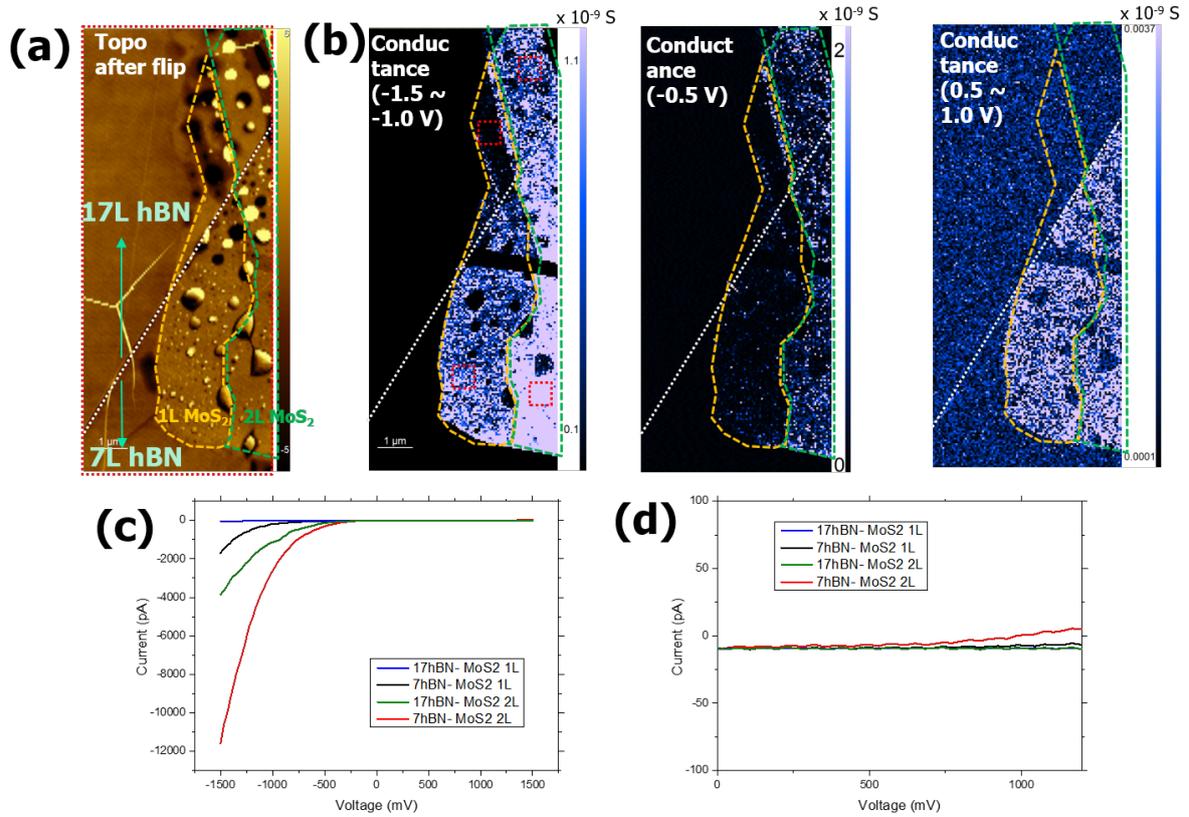

Figure S7. (a) Topography of MoS$_2$/hBN-Au buried interface (Case 3) with 7L & 17L hBN. 17L hBN is placed on top across the blue dotted line while below is 7L hBN. (b) Conductance map from -1.5 V to 1.0 V. (c) I-V Curves with each region as indicated in (b) by red square. (d) Magnified I-V curves at positive bias region.

**Tunneling current with respect to hBN thickness**

Tunneling current flowed when 5.6 nm *h*-BN was placed while did not at the presence of 14 nm *h*-BN on the range between 0.5 to 1.0 V. The voltage is much smaller than barrier height of *h*-BN, hence it is clearly result of tunneling. As far as know, 5.6 nm is the thickest barrier width reported in our best knowledge.

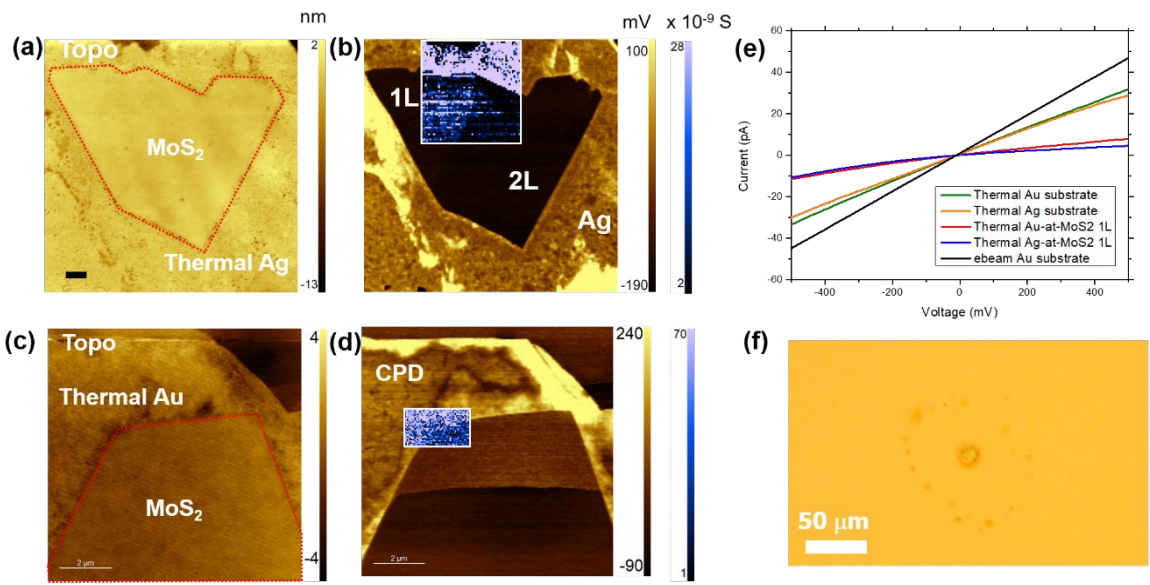

Figure S8. (a) Topography and (c) potential map of MoS$_2$-Ag buried interface by thermal evaporation. (a) Topography and (c) potential map of MoS$_2$-Au buried interface by thermal evaporation. The region enclosed by red dotted line is MoS$_2$ flake. (e) I-V curves comparison of Thermal Au and thermal Ag effect. (f) Au-assisted transferred MoS$_2$ sample made by sputtering.

**CPD and I-V curve analysis of MoS$_2$-Ag contact**

Ag layer was readily peeled off from the SiO2/Si substrate as it did for the Au Case 1 interface. Similar to Au Case 1 interface, potential of the Ag -MoS$_2$ 1L and 2L Case 1 interface were lower than Ag substrate and spatially uniform indicating reliable contact and Fermi level pinning (figure S8b). To be specific, the potential was also shifted by -183 mV for 1L region and -192 mV compared to Ag substrate. The I-V characteristics, however, exhibited much larger contact resistance than Au contact, recording 365 kΩ μm for 1L, although Ag work function is lower than that of Au. It indicates contact resistance is largely influenced by Fermi level pinning rather than metal work function. Interestingly, evaporation technique can be the major reason for the pinning.